\documentclass{article}
\usepackage[english]{babel}
\usepackage{amsfonts}    
\usepackage{amsmath}     
\usepackage{amssymb}     
\usepackage{amsthm}
\usepackage{latexsym}
\usepackage{graphicx}    
\usepackage{psfrag}      
\usepackage{color}       

\newtheorem{corollary}{Corollary}
\newtheorem{definition}{Definition}
\newtheorem{lemma}{Lemma}
\newtheorem{proposition}{Proposition}
\newtheorem{theorem}{Theorem}

\newcommand{\p}{\partial}

\newcommand{\sta}{\text{sta}}

\newcommand{\sss}{\scriptscriptstyle}
\newcommand{\Esss}{{\scriptscriptstyle E}}

\newcommand{\Psss}{{\scriptscriptstyle P}}

\newcommand{\pmsss}{{\scriptscriptstyle \pm}}

\newcommand{\RR}{\mathbb{R}}

\newcommand{\ZZ}{\mathbb{Z}}

\newcommand{\erm}{\mathrm{e}}
\newcommand{\irm}{\mathrm{i}}

\newcommand{\vrm}{\mathrm{v}}
\newcommand{\xrm}{\mathrm{x}}
\newcommand{\zrm}{\mathrm{z}}

\newcommand{\Ccal}{\mathcal{C}}
\newcommand{\Dcal}{\mathcal{D}}

\newcommand{\Jcal}{\mathcal{J}}

\newcommand{\Rcal}{\mathcal{R}}
\newcommand{\Scal}{\mathcal{S}}

\newcommand{\Zcal}{\mathcal{Z}}

\newcommand{\ba}{\begin{array}}
\newcommand{\ea}{\end{array}}
\newcommand{\bc}{\begin{corollary}}
\newcommand{\ec}{\end{corollary}}
\newcommand{\bd}{\begin{definition}}
\newcommand{\ed}{\end{definition}}
\newcommand{\be}{\begin{equation}}
\newcommand{\ee}{\end{equation}}
\newcommand{\bi}{\begin{itemize}}
\newcommand{\ei}{\end{itemize}}
\newcommand{\bl}{\begin{lemma}}
\newcommand{\el}{\end{lemma}}
\newcommand{\bp}{\begin{proposition}}
\newcommand{\ep}{\end{proposition}}
\newcommand{\bt}{\begin{theorem}}
\newcommand{\et}{\end{theorem}}
\newcommand{\bea}{\begin{eqnarray}}
\newcommand{\eea}{\end{eqnarray}}

\title{Renormalization and Quantum Scaling of
Frenkel--Kontorova Models}
\author{Nuno R. Catarino\footnote{Present address:
School of Mathematical and Computer Sciences,
Scott Russell Building, Heriot-Watt University,
Edinburgh EH14 4AS, U.K.}
\footnote{e-mail: catarino@ma.hw.ac.uk}
\&
Robert S. MacKay\footnote{e-mail: mackay@maths.warwick.ac.uk},\\
Mathematics Institute, University of Warwick \\
Gibbet Hill Road, Coventry CV4 7AL, U.K.}
\date{\today}

\begin{document}

\maketitle

\bibliographystyle{plain}


\abstract{
We generalise the classical Transition by Breaking of
Analyticity for the class of Frenkel--Kontorova models studied
by Aubry and others to non-zero Planck's constant and
temperature.
This analysis is based on the study of a renormalization
operator for the case of irrational mean spacing using
Feynman's functional integral approach.
We show how existing classical results extend to the quantum
regime.
In particular we extend MacKay's renormalization approach for
the classical statistical mechanics to deduce scaling of low
frequency effects and quantum effects.
Our approach extends the phenomenon of hierarchical melting
studied by Vallet, Schilling and Aubry to the quantum regime.


\section{Introduction}

The {\em Frenkel--Kontorova model} (FK) is a one-dimensional
lattice model exhibiting incommensurate structures.
It is a system of elastically coupled particles in an external
periodic potential (a discrete version of the sine--Gordon
model) with Lagrangian
\be
\label{eq:lagrangianGFK}
   L \left( x , \dot{x} \right) = \sum_{ n \in {\ZZ} } \left\{
                        \frac{ \dot{x}_n^2 }{2} - {\vrm}
                        \left( x_n , x_{ n + 1 } \right)
                        \right\} ,
\ee
where
\be
\label{eq:bondEnergyFK}
   \vrm \left( x, x^\prime \right) =
   \frac{1}{2} \left( x^\prime - x \right)^2
   + P \left( x^\prime - x \right)
   + \frac{u}{ 4 \pi^2 }\cos \left( 2 \pi x \right) ,
\ee
with constants $P$ (``pressure'' or $-P$ as tension) and $u$
(amplitude of the onsite potential).
We call successive pairs $\left( x_n , x_{ n + 1 } \right)$
{\em bonds} and $\vrm \left( x_n , x_{ n + 1 } \right)$
the {\em (potential) energy of the bond}.
Notice that in (\ref{eq:lagrangianGFK}) and
(\ref{eq:bondEnergyFK}) the particles' mass, the elastic
coupling strength and the period of the onsite potential are all
scaled to one.
The FK model is a particular case of the broader class of
{\em (Generalised) Frenkel--Kontorova models} (GFK) with
Lagrangian still given by~(\ref{eq:lagrangianGFK}) but the bond
action $\vrm$ is a generic $\Ccal^2$ function satisfying
\bea
\nonumber
   && \vrm \left( x + 1 , x^\prime + 1 \right) =
   \vrm \left( x , x^\prime \right) \\
\nonumber
   && \frac{ \p^2 \vrm }{ \p x \p x^\prime }
   \left( x , x^\prime \right) \leqslant - C < 0 \, .
\eea

In the space of parameters there are two
important limits.
In the {\em integrable} limit, the bond energy, $\vrm$, depends
only on $\left( x^\prime - x \right)$.
For the FK model (\ref{eq:bondEnergyFK}) the integrable limit
is attained at $u = 0$ and its {\em minimum energy
configurations} (i.e. $x \in {\RR}^{\ZZ}$ such that
$\forall M < N , V_{ M , N } \left( x \right) \equiv
\sum_{ n = M }^{ N - 1 } \vrm \left( x_n , x_{ n + 1 } \right)$
is minimum for all variations of $x_{ n }$ with fixed $x_M$ and
$x_N$) are arrays of equally spaced particles with
{\em mean spacing}
\be
\nonumber
   \rho := \lim_{ - M , N \rightarrow \infty } 
   \frac{ x_N - x_M }{ N - M }
\ee
simply $- P$.
In the {\em anti-integrable} limit \cite{art:AubryAbramovici90}
the onsite term dominates, which corresponds to
$u \rightarrow \infty$ in~(\ref{eq:bondEnergyFK}).
All particles are then in the minima of the potential and the
mean spacing is the closest integer to $- P$, or any value in
between the two if non-unique.

An interesting set of codimension-$2$ critical points occur
between these two regimes, often called {\em Transition by
Breaking of Analyticity} (TBA): in the space of parameters
$\left( u, P \right)$, for each irrational $\rho$ there is a
curve $P = P_\rho \left( u \right)$ of mean spacing $\rho$
containing a critical value $u_c$ (there may be more than one
$u_c$ depending on the potential).
The regime for $u$ less than $u_c$ is called {\em subcritical}
(or {\em sliding phase}) and above $u_c$ {\em supercritical}
(or {\em pinned phase})~\cite{art:AubryLeDaeron83}.
For the FK model with fixed mean spacing
$\rho = \gamma^{ - 1 }$, where
$\gamma = \left( 1 + \sqrt{5} \right) / 2$ is the golden mean,
the TBA is at the critical value
$u_c \simeq 0.971 \, 635 \, 406$~\cite{art:Greene79}.
This is the case most often studied in the literature, since it
is presumed to be the highest value of $u$ at which a TBA
occurs.

For the case of mean spacing $\rho = \gamma^{ - 1 }$,
relabelling the bonds $\vrm$ appropriately~\cite{art:MacKay91}
as $\tau$ and $\upsilon$ results in a Fibonacci sequence of
$\tau$s and $\upsilon$s where each $\upsilon$-bond is always
surrounded by $\tau$-bonds.
The TBA point can be viewed as a fixed point of the
renormalization operator that minimises the energy of the sum
of successive $\upsilon$ and $\tau$-bonds with suitably chosen
space and energy scalings, $\alpha, \Jcal \in {\RR}$
respectively depending on the pair
$\left( \upsilon , \tau \right)$ (actually it is also necessary
to subtract a constant and a quadratic coboundary but we will
suppress reference to these inessential
terms)~\cite{art:MacKay91}.
The renormalization operator has a nontrivial fixed
point\footnote{This has now been proved by
Koch~\cite{art:Koch04} by reformulation as a renormalization on
continuous--time Hamiltonian systems and rigorous
computer--assisted bounds.}
$\left( \bar{\upsilon} , \bar{\tau} \right)$ with
$\alpha \simeq - 1.414 \, 836 \, 0$ and
$\Jcal \simeq 4.399 \, 143 \, 9$.
This fixed point corresponds to the critical $u_c$ along the
curve in $\left( u , P \right)$ for $\rho = \gamma^{ - 1 }$, the
transition point between the subcritical and supercritical
regimes.
It has two unstable directions: one along the curve of constant
mean spacing, with eigenvalue $\delta \simeq 1.627 \, 950 \, 0$
and the other transverse to this curve in the
$\left( u , P \right)$ plane (which we'll call the $P$
direction), with eigenvalue
$\eta = - \Jcal / \gamma \simeq
- 2.681 \, 738 \, 4$~\cite{art:MacKay91,bok:MacKay93}.

Whereas the classical ground states of Frenkel--Kontorova models
have been extensively studied since the beginning of the 1980's
\cite{icl:Aubry82,art:Aubry83,art:Aubry83b,art:AubryLeDaeron83,art:PeyrardAubry83}
(see also the review in chapter~1 of~\cite{phd:Catarino04}, and
the book~\cite{bok:BraunKivshar04} for several aspects of the FK
model), the extension to the quantum regime of the classical
Transition by Breaking of Analyticity is still not fully
understood.
Most of the previous studies stem from the work of
Borgonovi~{\em et~al.}~\cite{art:BorgonoviEtAl89,art:BorgonoviEtAl90},
where the authors do a numerical study of FK model in the
supercritical region $u > u_c$ for the case of mean spacing
$\gamma^{ - 1 }$.
They introduce a `quantum hull function' for the expected
positions $\bar{x}_n$ as the extension of Aubry's hull function
(see~\cite{phd:Catarino04}, section~1.2) as well as the `quantum
$g$-function' (which reduces to
$\sin \left( 2 \pi \bar{x}_n \right)$ in the classical limit)
and verify that by increasing $\hslash$ the quantum hull
function becomes a smooth version of the classical hull function
and that $g_n$ tends to a sawtooth-like map.
Similar results have been later obtained using various numerical
or a combination of analytical and numerical methods
\cite{art:BermanEtAl94,art:BermanEtAl97,art:HuEtAl98,art:HoChou00}.

In a recent numerical study
Zhirov~{\em et al.}~\cite{art:ZhirovEtAl03}
claim to observe a `quantum phase transition' in the FK model
at a critical value of Planck's constant between `sliding phonon
gas' and a `pinned instanton glass'.
In fact we expect that due to KAM-type of arguments, for
sufficiently irrational mean spacing the phonon energy band will
survive for a small perturbation of the integrable limit $u = 0$,
where the interactions between phonons of different wavenumber
are small.
At the anti-integrable limit, on the other hand, when the
dominant interaction is the onsite periodic potential, the
quantum spectrum consists of $N$ degenerate sets of Bloch bands
($N$ being the total number of particles) whose width is due to
tunneling or instanton effects between distinct minima of the
onsite potential.
As $u$ is decreased from the anti-integrable limit, the
degeneracy between Bloch bands corresponding to distinct
particles should be lifted, widening the Bloch bands until they
merge into a unique phonon band for some non-zero value
$u_c \left( \hslash \right)$ (possibly not a unique curve).

In addition to the groundstates, it is physically significant
to study the effect of the TBA on the low temperature
statistical mechanics of FK models.
This was done for the classical case in \cite{art:MacKay95}.
The quantum statistical mechanics of FK models was considered in
a series of papers by Giachetti and Tognetti,
e.g. \cite{art:GiachettiTognetti85,art:GiachettiTognetti86}, but
we are not aware of any work on the effects of the TBA on the
quantum statistical mechanics.

The goal of this article is to study the transition above by
extending the minimum energy renormalization approach
in~\cite{art:MacKay91} to non-zero Planck's constant and
temperature, which is done in
section~\ref{sec:Renormalization_Q}.
This is done in a way similar to the classical non-zero
temperature extension performed in~\cite{art:MacKay95} but
requires first an extension of the classical case from ground
states to time-periodic solutions.
The strategy is to construct a renormalization operator,
$\Rcal$, which reduces to the ground state operator,
$\Rcal_\text{cgs}$, when $\hslash$ goes to zero, i.e.
$\left. \Rcal \right|_{ \hslash = 0 } = \Rcal_\text{cgs}$.
Then $\left( u , P , \dots , \hslash = 0 \right)$ is an
invariant subspace for $\Rcal$ which includes the critical fixed
point of renormalization corresponding to the TBA by
construction.

In the main section~\ref{sec:Renormalization_Q} we start by
defining a decimation procedure $\oplus$ for pairs of
``bond actions'' by doing the trace over intermediate particles
corresponding to a partial trace over the kernel and the
renormalization operator $\Rcal$ by composing the decimation
with the appropriate scalings.
The quantum eigenvalue $\kappa$ is also introduced.
Next, in subsection~\ref{ssec:SPAGSL_Q}, we show that the
classical ground state renormalization, $\Rcal_\text{cgs}$, is
obtained as the limit of zero Planck's constant and frequency of
quantum renormalization.
In subsection \ref{ssec:RenormalizationPhonons_Q} the value of
the quantum eigenvalue is determined by analysing the linearised
or phonon problem.
Finally in section \ref{ssec:TScaling_Q} the results
obtained are `Wick rotated' to obtain the scaling laws for
quantum thermal quantities, and an extension of the phenomenon
of hierarchical melting studied by Vallet, Schilling and Aubry
\cite{art:ValletEtAl86,art:SchillingAubry87,art:ValletEtAl88}
to the quantum regime is proposed.

\section{Renormalization and scaling}
\label{sec:Renormalization_Q}

We begin by introducing the renormalization procedure for the
action of time-periodic functions of prescribed period.
Consider the class of models with the following formal sum for
the action
\be
\nonumber
   \Scal \left( x ; T \right)
   = \sum_{ n \in \ZZ } s \left( x_n , x_{ n + 1 } ; T \right) .
\ee
for time--periodic functions $x_n$ of period $T$, where the
{\em bond action} $s$ is given by
\be
\label{eq:bondAction_Q}
  s \left( x , x^\prime ; T \right) =
    \int_0^T \left\{ \frac{ \dot{x}^2 }{2}
  - \vrm \left( x , x^\prime \right)
    \right\} d t ,
\ee
and $\vrm$ is a GFK bond potential energy.
The quantum renormalization will concern the trace of the kernel
which is
\be
\nonumber
   K^\prime \left( T , \hslash \right) =
   \int \Dcal^\prime x \,
   \erm^{ \frac{\irm}{\hslash} \Scal \left( x ; T \right) } ,
\ee
where the notation $\Dcal^\prime x$ means that the integration
is to be performed over the space of periodic paths $x$ with
period $T$ and includes the integration over the endpoints
$d x \left( 0 \right)$ (i.e.
$x \left( T \right) = x \left( 0 \right)$ and
$\Dcal^\prime x = d x \left( 0 \right) \, \Dcal x$).
Rescaling time $t \rightarrow t / T$ and defining
$\Omega = 2 \pi / T$ the bond action can be rewritten as
\be
\label{eq:bondActionRescaled_Q}
  s \left( x , x^\prime ; \frac{ 2 \pi }{\Omega} \right)
  = \frac{ 2 \pi }{\Omega}
    \int_0^1 \left\{ \frac{ \Omega^2 }{ 8 \pi^2 } \dot{x}^2
  - \vrm \left( x , x^\prime \right)
    \right\} d t ,
\ee
and now $s$ acts on the space of period-one functions
$x , x^\prime : \RR / \ZZ \rightarrow \RR$ or {\em loops}.

Now let $\xrm \in \RR^\ZZ$ be a classical ground state and call
a bond action $s$ a $\tau$ or an $\upsilon$ as for the
renormalization for classical ground states
in~\cite{art:MacKay91}.
For the case of mean spacing $\gamma^{ - 1 }$ the sequence of
bond actions then forms an infinite Fibonacci sequence of
$\tau$ and $\upsilon$ types of bonds where each $\upsilon$ is
always surrounded by $\tau$s.
At this point one wants to eliminate all particles $z$ from
sequences of the form $\upsilon \left( x , z ; \Omega \right)
+ \tau \left( z , x^\prime ; \Omega \right)$ (for simplicity
we use $\Omega$ as an argument of action bonds instead of
$2 \pi / \Omega$ from now on).
In order to do this define the following decimation operator
$\oplus$ acting on pairs of bond actions
$\left( \upsilon , \tau \right)$ as
\be
\label{eq:decimation_Q}
   \left( \upsilon \oplus \tau \right)
   \left( x , x^\prime ; \Omega \right) =
   - \irm \hslash \ln \int \Dcal^\prime z \,
   \erm^{ \frac{\irm}{\hslash} \left[
       \upsilon \left( x , z ; \Omega \right)
     + \tau \left( z , x^\prime ; \Omega \right) \right] } .
\ee
The decimated functional $\upsilon \oplus \tau$ still has the
form of an action bond in the sense that it acts on pairs of
loops $\left( x , x^\prime \right)$.
The renormalization operator is defined as the composition of a
decimation and scalings as
\be
\label{eq:renormalization_Q}
   \Rcal \left[
   \ba{c}
      \tau \left( x , x^\prime ; \Omega \right) \\
      \upsilon \left( x , x^\prime ; \Omega \right)
   \ea
   \right] = \frac{\Jcal}{\varepsilon} \left[
   \ba{c}
      \left( \upsilon \oplus \tau \right)
      \left( x / \alpha, x^\prime / \alpha ;
             \Omega / \varepsilon \right) \\
      \tau \left( x / \alpha, x^\prime / \alpha ;
             \Omega / \varepsilon \right)
   \ea
   \right] ,
\ee
which includes a scaling of frequencies, $\varepsilon$, still
undetermined at this point.
The global scaling of bond actions is $\Jcal / \varepsilon$
instead of just $\Jcal$ because the renormalization now acts on
actions instead of energies as in the classical ground state
case.
By performing the composition of (\ref{eq:renormalization_Q})
and (\ref{eq:decimation_Q}) explicitly one can see that the
natural scale factor $\kappa = \Jcal / \varepsilon$ for Planck's
constant arises and $\Rcal$ can be seen as acting on the
extended space of
$\left( \tau , \upsilon , \hslash \right)$ for $\hslash \geq 0$
as
\be
\label{eq:renormalizationExtended_Q}
   \Rcal:
   \left\{
   \ba{l}
      \tilde{\tau} \left( x , x^\prime ; \Omega \right)
      = - \irm \kappa \hslash \int \Dcal^\prime z \,
      \erm^{ \frac{\irm}{ \kappa \hslash } \frac{\Jcal}{\varepsilon}
      \left\{
        \upsilon \left( x / \alpha , z
                      ; \Omega /\varepsilon \right)
      + \tau \left( z , x^\prime / \alpha
                  ; \Omega / \varepsilon \right)
      \right\} } \\
      \tilde{\upsilon} \left( x , x^\prime ; \Omega \right)
      = \frac{\Jcal}{\varepsilon}
      \tau \left( x / \alpha , x^\prime / \alpha
                ; \Omega / \varepsilon \right) \\
      \tilde{\hslash} = \kappa \hslash
   \ea \right. .
\ee
We interpret $\kappa$ as the eigenvalue of $\Rcal$ in the
direction of Planck's constant.

\subsection{Semiclassical approximation and ground state limit}
\label{ssec:SPAGSL_Q}

As a first step to connect the quantum renormalization
(\ref{eq:renormalization_Q}) with the ground state
renormalization of~\cite{art:MacKay91}, one can look at the
decimation operator (\ref{eq:decimation_Q}) in the stationary
phase approximation for small $\hslash$:
\bea
\nonumber
   \left( \upsilon \oplus \tau \right)
   \left( x , x^\prime ; \Omega \right)
   = - \irm \hslash \ln \int \Dcal^\prime z \,
   \delta \left( z - z_\text{cl.}
   \left( x , x^\prime ; \Omega \right) \right) \times \\
\label{eq:decimationSPA1_Q}
   \times \erm^{ \frac{\irm}{\hslash}
   \left\{ \upsilon \left( x , z ; \Omega \right)
         + \tau \left( z , x^\prime ; \Omega \right)
   \right\} } .
\eea
Here $z_\text{cl.} \left( x , x^\prime ; \Omega\right)$ is
the classical path of period one satisfying
the Euler--Lagrange equations
\be
\label{eq:EulerLagrange_Q}
   \left. \frac{ \delta
   \left[ \upsilon \left( x , z ; \Omega \right)
        + \tau \left( z , x^\prime ; \Omega \right) \right]
             }{ \delta z } \right|_{ z = z_\text{cl.}
                         \left( x , x^\prime ; \Omega \right) }
   = 0 .
\ee
In the classical limit one is thus left with a dynamical problem
(\ref{eq:EulerLagrange_Q}) and the decimated action
(\ref{eq:decimationSPA1_Q}) can be rewritten as
\be
\label{eq:decimationSPA2_Q}
   \mathop{\sta}_{ z \in \text{loops} }
   \left[ \upsilon \left( x , z ; \Omega \right)
        + \tau \left( z , x^\prime ; \Omega \right) \right] ,
\ee
the sum of bond actions evaluated at $z = z_\text{cl.}$, given
by~(\ref{eq:EulerLagrange_Q}), which stationarises the sum
$\upsilon + \tau$ over the space of all loops
$\left\{ z : \RR / \ZZ \rightarrow \RR \right\}$.

\subsubsection{Ground state limit}

The ground state decimation can now be taken as the limit
$\Omega \rightarrow 0$ of (\ref{eq:decimationSPA1_Q}) or
(\ref{eq:decimationSPA2_Q}).
To see this, notice that in this limit only the classical ground
states contribute to the kernel and
\be
\nonumber
   \int \Dcal^\prime z \,
   \erm^{ \frac{\irm}{\hslash} \left\{
          \upsilon \left( x , z \right) 
        + \tau \left( z , x^\prime \right) \right\} }
   \simeq \erm^{ - \frac{ 2 \pi \irm }{ \hslash \Omega } \left\{
          \vrm^{ ( \upsilon ) }
          \left( \xrm , \zrm \right) 
        + \vrm^{ ( \tau ) }
          \left( \zrm , \xrm^\prime \right) \right\} } ,
\ee
where $\left( \xrm , \zrm , \xrm^\prime \right)$ is a segment of
a classical ground state and
$\zrm \left( \xrm , \xrm^\prime \right)$ minimises the sum
$\upsilon + \tau$.
Taking the logarithm and multiplying by $- \irm \hslash$, this
results in the decimation (\ref{eq:decimation_Q}) being
\be
\label{eq:decimationCGS_Q}
   \left( \upsilon \oplus \tau \right)
   \left( x , x^\prime ; 0 \right)
   = \lim_{ \Omega \rightarrow 0 }
   - \frac{ 2 \pi }{\Omega} \min_{ \zrm \in \RR }
   \left[ \vrm^{ ( \upsilon ) } \left( \xrm , \zrm \right)
        + \vrm^{ ( \tau ) } \left( \zrm , \xrm^\prime \right)
   \right]
\ee
(the factor $2 \pi / \Omega$ is kept in the above expression
simply to control the divergence of decimation).
Apart from the limit factor
$\lim_{ \Omega \rightarrow 0 } - 2 \pi / \Omega$ this is in fact
the classical ground state decimation in~\cite{art:MacKay91} and
so the renormalization (\ref{eq:renormalization_Q}) becomes
\bea
\nonumber
   \Rcal \left[
   \ba{c}
      \tau \left( x , x^\prime ; 0 \right) \\
      \upsilon \left( x , x^\prime ; 0 \right)
   \ea \right]
   &\simeq& \lim_{ \Omega \rightarrow 0 } \left[
   \ba{c}
      - \frac{ 2 \pi }{\Omega}
      \Jcal \left( \vrm^{ ( \upsilon ) } \oplus_\text{cgs}
               \vrm^{ ( \tau ) } \right)
      \left( \xrm / \alpha, \xrm^\prime / \alpha \right) \\
      - \frac{ 2 \pi }{\Omega} 
      \Jcal \vrm^{ ( \tau ) }
        \left( \xrm / \alpha, \xrm^\prime / \alpha \right)
   \ea \right] \\
\label{eq:renormalizationQtoGS_Q}
   &=& \lim_{ \Omega \rightarrow 0 }
     - \frac{ 2 \pi }{\Omega}
   \Rcal_\text{cgs} \left[
   \ba{c}
      \vrm^{ ( \tau ) }
      \left( \xrm , \xrm^\prime \right) \\
      \vrm^{ ( \upsilon ) }
      \left( \xrm , \xrm^\prime \right)
   \ea
   \right] ,
\eea
where $\oplus_\text{cgs}$ and $\Rcal_\text{cgs}$ are the
decimation and renormalization operators for the classical
ground states, and $\varepsilon \in \RR$ is still undetermined
at this point.
The renormalization $\Rcal$ has therefore a fixed point (the TBA
fixed point), with ground states of mean spacing
$\rho = \gamma^{ - 1 }$, in the $\Omega = 0$ subspace consisting
of
\be
\nonumber
   \left[
   \ba{c}
      \bar{\tau} \left( x , x^\prime ; 0 \right) \\
      \bar{\upsilon} \left( x , x^\prime ; 0 \right)
   \ea
   \right] = \lim_{ \Omega \rightarrow 0 }
   - \frac{ 2 \pi }{\Omega}
   \left[
   \ba{c}
      \bar{\vrm}^{ ( \tau ) } \left( x , x^\prime \right) \\
      \bar{\vrm}^{ ( \upsilon ) } \left( x , x^\prime \right)
   \ea
   \right] ,
\ee
where $\left( \bar{\vrm}^\tau , \bar{\vrm}^\upsilon \right)$ is
the fixed point of the ground state renormalization operator
$\Rcal_\text{cgs}$, because applying
expression~(\ref{eq:renormalizationQtoGS_Q}) at the TBA fixed
point results in
\be
\nonumber
   \Rcal \left[
   \ba{c}
      \bar{\tau} \left( x , x^\prime ; 0 \right) \\
      \bar{\upsilon} \left( x , x^\prime ; 0 \right)
   \ea
   \right] = \lim_{ \Omega \rightarrow 0 }
   - \frac{ 2 \pi }{\Omega} \left[
   \ba{c}
      \bar{\vrm}^{ ( \tau ) }
      \left( \xrm , \xrm^\prime \right) \\
      \bar{\vrm}^{ ( \upsilon ) }
      \left( \xrm , \xrm^\prime \right)
   \ea
   \right] = \left[
   \ba{c}
      \bar{\tau} \left( x , x^\prime ; 0 \right) \\
      \bar{\upsilon} \left( x , x^\prime ; 0 \right)
   \ea
   \right] .
\ee

\subsection{Renormalization for the phonon spectrum}
\label{ssec:RenormalizationPhonons_Q}

Close to the ground state, for small $\Omega^2$, the relevant
contributions are approximately given by the normal modes or
{\em phonons}.
For a GFK model at irrational mean spacing ground state, $\rho$,
the phonon spectrum includes zero in the subcritical regime, but
the minimum frequency or {\em phonon gap} is positive above the
critical point $u_c$~\cite{art:AubryLeDaeron83}.
The phonon contribution to the trace of the kernel is given by
doing a quadratic approximation which is
(with $x_n = \xrm_n + \xi_n$)
\bea
\label{eq:traceKPhonon_Q}
   K^\prime \left( \xrm , \frac{ 2 \pi }{\Omega} \right)
   &\simeq&
   \erm^{ - \frac{ 2 \pi \irm }{ \hslash \Omega }
   \sum_n
   \vrm^{ ( n ) } \left( \xrm_n , \xrm_{ n + 1 } \right) }
   \times \\
\nonumber
   &\times& \int \Dcal^\prime \xi \,
   \erm^{ \frac{ \pi \irm }{ \hslash \Omega }
   \sum_n \int_0^1 \left\{
   \frac{\Omega^2}{ 4 \pi^2 } m^{ ( n ) }
   \left( \dot{\xi}_n , \dot{\xi}_{ n + 1 } \right)
   - u^{ ( n ) }
   \left( \xi_n , \xi_{ n + 1 } \right) \right\} d t } ,
\eea
where the kinetic term was generalised to a symmetric quadratic
form in the velocities
\be
\nonumber
   m^{ ( n ) } \left( \dot{\xi} , \dot{\xi}^\prime \right)
   = m_{ \sss 1 1 }^{ ( n ) } \dot{\xi}^2
   + 2 m_{ \sss 1 2 }^{ ( n ) } \dot{\xi} \, \dot{\xi}^\prime
   + m_{ \sss 2 2 }^{ ( n ) } \left. {\dot{\xi}}^\prime \right.^2
\ee
(as will become clear later on the renormalization operator
introduces coupling between the velocities of neighbouring
particles), and $u$ is also a symmetric quadratic form
\be
\nonumber
   u^{ ( n ) } \left( \xi , \xi^\prime \right)
   = u_{ \sss 2 2 }^{ ( n ) } \xi^2
   - 2 u_{ \sss 1 2 }^{ ( n ) } \xi \xi^\prime
   + u_{ \sss 1 1 }^{ ( n ) } { \xi^\prime }^2 .
\ee

The first term in (\ref{eq:traceKPhonon_Q}) is simply the
classical ground state and corresponds to
(\ref{eq:renormalizationQtoGS_Q}), so one wants to apply the
decimation (\ref{eq:decimation_Q}) to the sum of the quadratic
parts of bond actions of the form
\be
\nonumber
   \frac{\pi}{\Omega} \left\{
   \frac{\Omega^2}{ 4 \pi^2 }
     \left[ m^{ ( \upsilon ) }
            \left( \dot{\xi} , \dot{\zeta} \right)
          + m^{ ( \tau ) }
            \left( \dot{\zeta} , \dot{\xi}^\prime \right)
   \right]
   - \left[ u^{ ( \upsilon ) } \left( \xi , \zeta \right)
          + u^{ ( \tau ) } \left( \zeta , \xi^\prime \right)
   \right]
   \right\} .
\ee
Because this sum is quadratic in the particle to eliminate,
$\zeta$, the corresponding functional integral can be calculated
(for example by Gaussian integration of the Fourier transformed
sum of action bonds~\cite{bok:ZinnJustin02}) and the
semiclassical approximation is exact in this case.
To first order in $\Omega^2$ the result is\footnote{There is
possibly also a logarithmic term due to the integration measure
which corresponds to a redefinition of the ground state energy.}
\bea
\nonumber
   \erm^{ \frac{ \irm \pi }{ \hslash \Omega }
   \left\{ \frac{ \Omega^2 }{ 4 \pi^2 }
     \left[
       m^{ ( \upsilon ) }
       \left( \dot{\xi} , \dot{\zeta} \right)
     + m^{ ( \tau ) }
       \left( \dot{\zeta} , \dot{\xi}^\prime \right)
     \right]
   - \left[ u^{ ( \upsilon ) } \left( \xi , \zeta \right)
          + u^{ ( \tau ) } \left( \zeta , \xi^\prime \right)
     \right]
   \right\} } \simeq \\
\nonumber
   \simeq \erm^{ \frac{ \irm \pi }{ \hslash \Omega }
   \left\{ \frac{ \Omega^2 }{ 4 \pi^2 }
     \tilde{m}^{ ( \tau ) }
     \left( \dot{\xi} , \dot{\xi}^\prime \right)
   - \tilde{u}^{ ( \tau ) }
     \left( \xi , \xi^\prime \right)
   \right\} } ,
\eea
where $\tilde{m}^{ ( \tau ) }$ is a new symmetric quadratic form
with components given by
\bea
\nonumber
   \tilde{m}_{ \sss 1 1 }^{ ( \tau ) }
   &=& m_{ \sss 1 1 }^{ ( \upsilon ) }
     + \frac{ 2 m_{ \sss 1 2 }^{ ( \upsilon ) }
                u_{ \sss 1 2 }^{ ( \upsilon ) }
           }{ u_{ \sss 2 2 }^{ ( \upsilon ) }
            + u_{ \sss 1 1 }^{ ( \tau ) } }
     + \frac{ \left( m_{ \sss 2 2 }^{ ( \upsilon ) }
                   + m_{ \sss 1 1 }^{ ( \tau ) } \right)
              { u_{ \sss 1 2 }^{ ( \upsilon ) } }^2
           }{ \left( u_{ \sss 2 2 }^{ ( \upsilon ) }
                   + u_{ \sss 1 1 }^{ ( \tau ) }
              \right)^2 } \\
\label{eq:barMTau_Q}
   \tilde{m}_{ \sss 1 2 }^{ ( \tau ) }
   &=& \frac{ m_{ \sss 1 2 }^{ ( \upsilon ) }
              u_{ \sss 1 2 }^{ ( \tau ) }
            + m_{ \sss 1 2 }^{ ( \tau ) }
              u_{ \sss 1 2 }^{ ( \upsilon ) }
           }{ u_{ \sss 2 2 }^{ ( \upsilon ) }
            + u_{ \sss 1 1 }^{ ( \tau ) } }
     + \frac{ \left( m_{ \sss 2 2 }^{ ( \upsilon ) }
                   + m_{ \sss 1 1 }^{ ( \tau ) } \right)
              u_{ \sss 1 2 }^{ ( \upsilon ) }
              u_{ \sss 1 2 }^{ ( \tau ) }
           }{ \left( u_{ \sss 2 2 }^{ ( \upsilon ) }
                   + u_{ \sss 1 1 }^{ ( \tau ) } \right)^2 } \\
\nonumber
   \tilde{m}_{ \sss 2 2 }^{ ( \tau ) }
   &=& m_{ \sss 2 2 }^{ ( \tau ) }
     + \frac{ 2 m_{ \sss 1 2 }^{ ( \tau ) }
                u_{ \sss 1 2 }^{ ( \tau ) }
           }{ u_{ \sss 2 2 }^{ ( \upsilon ) }
            + u_{ \sss 1 1 }^{ ( \tau ) } }
     + \frac{ \left( m_{ \sss 2 2 }^{ ( \upsilon ) }
                   + m_{ \sss 1 1 }^{ ( \tau ) } \right)
              { u_{ \sss 1 2 }^{ ( \tau ) } }^2 
           }{ \left( u_{ \sss 2 2 }^{ ( \upsilon ) }
                   + u_{ \sss 1 1 }^{ ( \tau ) } \right)^2 }
\eea
and the new form $\tilde{u}^{ ( \tau ) }$ has components
\bea
\nonumber
   \tilde{u}_{ \sss 1 1 }^{ ( \tau ) }
   &=& u_{ \sss 1 1 }^{ ( \upsilon ) }
     - \frac{ { u_{ \sss 1 2 }^{ ( \upsilon ) } }^2
             }{ u_{ \sss 2 2 }^{ ( \upsilon ) }
              + u_{ \sss 1 1 }^{ ( \tau ) } } \\
\label{eq:barUTau_Q}
   \tilde{u}_{ \sss 1 2 }^{ ( \tau ) }
   &=& \frac{ u_{ \sss 1 2 }^{ ( \upsilon ) }
              u_{ \sss 1 2 }^{ ( \tau ) }
           }{ u_{ \sss 2 2 }^{ ( \upsilon ) }
            + u_{ \sss 1 1 }^{ ( \tau ) } } \\
\nonumber
   \tilde{u}_{ \sss 2 2 }^{ ( \tau ) }
   &=& u_{ \sss 2 2 }^{ ( \tau ) }
            - \frac{ { u_{ \sss 1 2 }^{ ( \tau ) } }^2
                 }{ u_{ \sss 2 2 }^{ ( \upsilon ) }
                  + u_{ \sss 1 1 }^{ ( \tau ) } } .
\eea
With these new quadratic forms, the decimation
(\ref{eq:decimation_Q}) becomes simply
\bea
\label{eq:decimationPhonon_Q}
   \left( \upsilon \oplus \tau \right)
   \left( x , x^\prime ; \Omega \right)
   &=&\text{(c.g.s. decimation)} + \\
\nonumber
   &+&\frac{1}{\Omega} \int_0^1
     \Omega^2 \tilde{m}^{ ( \tau ) }
     \left( \xi , \xi^\prime \right)
   - \tilde{u}^{ ( \tau ) }
     \left( \xi , \xi^\prime \right) \, d t ,
\eea
where `c.g.s decimation' is the decimation
(\ref{eq:decimationCGS_Q}).
For the renormalization (\ref{eq:renormalization_Q}) one also
needs the `undecimated' bond actions which correspond to
isolated bond actions of type $\tau$, so one should define also
\be
\label{eq:barUMUpsilon_Q}
   \ba{l}
      \tilde{m}^{ ( \upsilon ) } \left( \xi , \xi^\prime \right)
      = m^{ ( \tau ) } \left( \xi , \xi^\prime \right) \\
      \tilde{u}^{ ( \upsilon ) } \left( \xi , \xi^\prime \right)
      = u^{ ( \tau ) } \left( \xi , \xi^\prime \right) .
   \ea
\ee
For a GFK model (\ref{eq:bondAction_Q}) at the start of
renormalization, for either type $\tau$ or
$\upsilon$ of bonds the quadratic forms are
$m_{ \sss 1 2 } = 0$,
$m_{ \sss 1 1 } + m_{ \sss 2 2 } = 1$,
$u_{ \sss j j } = \vrm_{ \sss , j j }
\left( \xrm , \xrm^\prime \right)$
for $j = 1 , 2$ and $u_{ \sss 1 2 } =
- \vrm_{ \sss , 1 2 } \left( \xrm , \xrm^\prime \right)$,
where the subscripts in $\vrm$ denote differentiation with
respect to the first and second variables and
$\left( \xrm , \xrm^\prime \right)$ is a segment of a classical
ground state.
Because $u$ is then dependent on
$\left( \xrm , \xrm^\prime \right)$, after iterating the above
transformations (\ref{eq:barMTau_Q}), (\ref{eq:barUTau_Q}) and
(\ref{eq:barUMUpsilon_Q}) one ends up with a set of asymptotic
quadratic forms depending on the ground
state, $\bar{m}^{ ( \tau ) }_{ \xrm , \xrm^\prime }$
$\bar{m}^{ ( \upsilon ) }_{ \xrm , \xrm^\prime }$,
$\bar{u}^{ ( \tau ) }_{ \xrm , \xrm^\prime }$ and
$\bar{u}^{ ( \upsilon ) }_{ \xrm , \xrm^\prime }$ which scale
by factors\footnote{\label{footnote:scaling_Q}
In~\cite{art:MacKay91} the scaling is actually for the
quantities
$a_n := u_{ \sss 2 2 }^{ ( n - 1 ) } + u_{ \sss 1 1 }^{ ( n ) }$,
$b_n := u_{ \sss 1 2 }^{ ( n ) }$,
$c_n :=
m_{ \sss 2 2 }^{ ( n - 1 ) } + m_{ \sss 1 1 }^{ ( n ) }$
and $d_n := m_{ \sss 1 2 }^{ ( n ) }$ with scaling constants
$\omega \simeq 1.255 \, 071$ for $a_n$ and $b_n$ and
$\beta / \alpha$ for $c_n$ and $d_n$, so the first and third
equations in (\ref{eq:scalingsMU_Q}) are defined up to a
quadratic coboundary.
Here we use the scale factors for frequency $\varepsilon :=
\sqrt{ \omega \Jcal / \alpha^2 } \simeq 1.649 \, 415$
and energy $\Jcal = \alpha \beta$ instead.
The origin of $\omega$ is still a mystery.}
$\omega =\alpha^2 \varepsilon^2 / \Jcal \simeq 1.255 \, 071$ for
the mass forms and $\Jcal / \alpha^2$ for the potential
forms~\cite{art:MacKay91}, i.e.
\be
\label{eq:scalingsMU_Q}
   \left\{
   \ba{l}
      \tilde{\bar{m}}_{ \xrm , \xrm^\prime }^{ ( \tau ) }
      \simeq
      \frac{ \alpha^2 \varepsilon^2 }{\Jcal}
      \bar{m}_{ \xrm , \xrm^\prime }^{ ( \tau ) } \\
      \tilde{\bar{m}}_{ \xrm , \xrm^\prime }^{ ( \upsilon ) }
      \simeq
      \frac{ \alpha^2 \varepsilon^2 }{\Jcal}
      \bar{m}_{ \xrm , \xrm^\prime }^{ ( \tau ) } \\
      \tilde{\bar{u}}_{ \xrm , \xrm^\prime }^{ ( \tau ) }
      \simeq
      \frac{\alpha^2}{\Jcal}
      \bar{u}_{ \xrm , \xrm^\prime }^{ ( \tau ) } \\
      \tilde{\bar{u}}_{ \xrm , \xrm^\prime }^{ ( \upsilon ) }
      \simeq
      \frac{\alpha^2}{\Jcal}
      \bar{u}_{ \xrm , \xrm^\prime }^{ ( \upsilon ) }
   \ea
   \right. .
\ee
Finally, using (\ref{eq:barMTau_Q}--\ref{eq:scalingsMU_Q}), and
defining
\bea
\nonumber
   \bar{\tau} \left( x , x^\prime ; \Omega\right) &:=&
   \frac{\pi}{\Omega} \left[
   - \bar{\vrm}^{ ( \tau ) }
     \left( \xrm , \xrm^\prime \right) \right. \\
\nonumber
   &+& \left. \int_0^1 \left\{ \frac{\Omega^2}{ 4 \pi^2 }
     \bar{m}_{ \xrm , \xrm^\prime }^{ ( \tau ) }
     \left( \xi , \xi^\prime \right)
   - \bar{u}_{ \xrm , \xrm^\prime }^{ ( \tau ) }
     \left( \xi , \xi^\prime \right)  \right\} d t \right] \\
\nonumber
   \bar{\upsilon} \left( x , x^\prime ; \Omega \right) &:=&
   \frac{\pi}{\Omega} \left[
   - \bar{\vrm}^{ ( \upsilon ) }
     \left( \xrm , \xrm^\prime \right) \right. \\
\nonumber
   &+& \left. \int_0^1 \left\{ \frac{\Omega^2}{ 4 \pi^2 }
     \bar{m}_{ \xrm , \xrm^\prime }^{ ( \tau ) }
     \left( \xi , \xi^\prime \right)
   - \bar{u}_{ \xrm , \xrm^\prime }^{ ( \tau ) }
     \left( \xi , \xi^\prime \right) \right\} d t \right] ,
\eea
the renormalization (\ref{eq:renormalizationExtended_Q}) at
$\left( \bar{\tau} , \bar{\upsilon} , \hslash \right)$ becomes
\bea
\nonumber
   \Rcal \left[
   \ba{c}
      \bar{\tau} \left( x , x^\prime ; \Omega \right) \\
      \bar{\upsilon} \left( x , x^\prime ; \Omega \right) \\
      \hslash
   \ea
   \right]
\nonumber
   &\simeq& \left[
   \ba{c}
      \bar{\tau} \left( x , x^\prime ; \Omega \right) \\
      \bar{\upsilon} \left( x , x^\prime ; \Omega \right) \\
      \kappa \hslash
   \ea
   \right] .
\eea
Thus, by including the scaling of frequencies
$\varepsilon \simeq 1.649 \, 415$ (see footnote
\ref{footnote:scaling_Q}) in the renormalization
(\ref{eq:renormalization_Q}), the point
$\left( \bar{\tau} , \bar{\upsilon} , 0 \right)$ becomes an
approximate fixed point for small $\Omega$
(actually a line of fixed points parametrised by $\Omega$).
In particular the phonon spectrum is asymptotically self-similar
under scaling by $\varepsilon$ in the direction of $\Omega$.

Finally, note that fixing $\varepsilon$ also determines the
eigenvalue in the $\hslash$-direction\footnote{This result was
published in \cite{art:CatarinoMacKay03} containing a mistake:
$k = \Jcal \varepsilon$ instead of the correct value
$\kappa = \Jcal / \varepsilon$.}
as $\kappa = \Jcal / \varepsilon \simeq 2.630 \, 716$.

\subsection{Scaling of the kernel}
\label{ssec:KScaling_Q}

The results above imply that the effect of including both the
frequencies and quantum directions, close to the TBA fixed point
(for $\Delta u := u - u_c$, $\Delta P :=
P - P_{ \gamma^{ - 1 } } \left( u_c \right)$, $\Omega$
and $\hslash$ small), is that the following asymptotic relation
of bond actions (regarded as functions in parameter space) holds
\be
\label{eq:scalingEquivRel_Q}
   \left( \tau , \upsilon \right)
   \left( \Delta u
        , \Delta P
        , \Omega
        , \hslash \right) \simeq
   \frac{\Jcal}{\varepsilon}
   \left(  \tau , \upsilon \right)
   \left( \delta \Delta u , \eta \Delta P
        , \varepsilon \Omega , \kappa \hslash \right) .
\ee
If $K_{ J , F_m }^\prime$ is the trace of the kernel for a chain
of size $F_m$, the $m$th Fibonacci number, in the discretised
form with $J$ `time steps' (such that $K_{ F_m }^\prime =
\lim_{ J \rightarrow \infty } K^\prime_{ J , F_m }$), then
\be
\nonumber
   K_{ J , F_m }^\prime
   \left( \Delta u
        , \Delta P
        , \Omega
        , \hslash \right)
   \simeq
   K_{ J , F_{ m - 1 } }^\prime
   \left( \delta \, \Delta u
        , \eta \, \Delta P
        , \varepsilon \, \Omega
        , \kappa \, \hslash \right)
   \left( \frac{ \sqrt{\Jcal}
              }{ \left| \alpha \right| \varepsilon }
   \right)^{ J F_{ m - 1 } } .
\ee
Here the multiplying factor comes from the functional
integration measure due to the change of coordinates (here with
diagonal mass components
$\mu^{ ( n ) } = \bar{m}^{ ( n - 1 ) }_{ \sss 2 2 } +
\bar{m}^{ ( n ) }_{ \sss 1 1 }$)
\bea
\nonumber
   \prod_n \left( \Dcal_J^\prime \frac{x_n}{\alpha} \right)
   \left( \Omega , \hslash \right)
   &=& \prod_n \prod_{ j = 0 }^{ J - 1 }
   \sqrt{ \frac{ \mu^{ ( n ) } \Omega J
              }{ 4 \pi^2 \irm \, \hslash } } \,
   \frac{ d x^{ ( j ) }_n }{ \left| \alpha \right| } \\
\nonumber
   &=& \prod_n \left( \frac{ \sqrt{\Jcal}
                  }{ \left| \alpha \right| \varepsilon }
       \right)^J
\left( \Dcal_J^\prime \tilde{x}_n \right)
\left( \varepsilon \Omega , \kappa \hslash
\right) ,
\eea
where $\tilde{x}_n$ are the positions of the renormalized
particles.


\subsection{Non-zero temperature scaling}
\label{ssec:TScaling_Q}

The quantum partition function at temperature $\Theta$ can be
easily obtained from the trace of the kernel by Wick rotation,
i.e. doing $2 \pi / \Omega = - \irm \beta \hslash$, where
$\beta = 1 / \Theta$.
The quantum partition function is then
\be
\nonumber
   \Zcal \left( \beta \hslash , \hslash \right)
   = K^\prime \left( - \irm \beta \hslash , \hslash \right)
   = \int \Dcal^\prime x \,
   \erm^{ - \beta \Scal_E \left( x; \beta \hslash \right) },
\ee
where the {\em Euclidean action} for the FK model is
\bea
\nonumber
   \Scal_E &=& \sum_n s_{ \sss E }
   \left( x_n, x_{ n + 1 } ; \beta \hslash \right) \\
\nonumber
   &=& \sum_n \int_0^1 \left\{
   \frac{ \dot{x}_n^2 }{ 2 \left( \beta \hslash \right)^2 }
   + \vrm \left( x_n, x_{ n + 1 } \right)  \right\}
   d t_{ \Esss } ,
\eea
and the integration is now over the rescaled Euclidean time
$t_\Esss \in \RR / \ZZ$.
Comparing with (\ref{eq:bondActionRescaled_Q}), the {\em Euclidean
bond action} $s_\Esss$ can be written as
\be
\nonumber
   s_\Esss \left( x , x^\prime ; \beta \hslash \right)
   = \left[ - \frac{\Omega}{ 2 \pi } \,
   s \left( x , x^\prime ; \Omega \right)
   \right]_{ \Omega = 2 \pi \irm / \beta \hslash} ,
\ee
and has units of an energy instead of an action because of the
multiplying factor $\Omega / 2 \pi$.
This gives another direction to which renormalization can be
extended, the temperature direction $\Theta = 1 / \beta$.
Define then the operator $\oplus_\Esss$ by `Wick rotating' the
operator $\oplus$ for the trace of the
kernel~(\ref{eq:decimation_Q}), to obtain
\be
\nonumber
   \left( \upsilon_\Esss \oplus_\Esss \tau_\Esss \right)
   \left( x , x^\prime ; \beta \hslash \right) =
   - \beta^{ - 1 } \ln \int \Dcal^\prime z \,
   \erm^{ - \beta \left[
       \upsilon_\Esss \left( x , z ; \beta \hslash \right)
   + \tau_\Esss
     \left( z , x^\prime ; \beta \hslash \right) \right] } ,
\ee
and the renormalization as
\be
\nonumber
   \Rcal_\Esss \left[
   \ba{c}
      \tau_\Esss
      \left( x , x^\prime ; \beta \hslash \right) \\
      \upsilon_\Esss
      \left( x , x^\prime ; \beta \hslash \right)
   \ea
   \right] = \left[
   \ba{c}
      \Jcal \tau_\Esss
      \left( x / \alpha , x^\prime / \alpha
           ; \varepsilon \beta \hslash \right) \\
      \Jcal \upsilon_\Esss
      \left( x / \alpha , x^\prime / \alpha
           ; \varepsilon \beta \hslash \right)
   \ea \right] .
\ee
Therefore, close to the TBA fixed point the renormalization
operator leads to the corresponding relation to
(\ref{eq:scalingEquivRel_Q}) for Euclidean bond actions:
\be
\nonumber
   \left( \tau_\Esss , \upsilon_\Esss \right)
   \left( \Delta u , \Delta P , \beta \hslash , \hslash \right)
   \simeq \Jcal \left( \tau_\Esss , \upsilon_\Esss \right)
   \left( \delta \Delta u , \eta \Delta P
        , \varepsilon^{ \sss - 1 } \beta \hslash
        , \kappa \hslash \right) ,
\ee
so under the renormalization $\Rcal_\Esss$ there is a fixed
point at $\hslash = 0$, $\beta \hslash = \infty$, with an
unstable eigenvalue of $\kappa \varepsilon = \Jcal \simeq 4.339
\, 143 \, 9$ in the temperature direction.
Although this agrees with the scaling of \cite{art:MacKay95} for
classical statistical mechanics, the result here includes the
momentum contribution and corresponds to the quantum correction
of the low temperature classical result, in the region
$\Theta \ll \hslash$ in which the classical partition function
is not valid (see \cite{phd:Catarino04}, section~3.2).
Denoting by
$\Zcal_{ F_m } \left( \Delta u , \Delta P , \Theta , \hslash \right)$
the quantum partition function for a chain of $F_m$ particles
at temperature $\Theta$, regarded as a function in parameter space,
the renormalization picture leads to
\be
\nonumber
   \Zcal_{ F_m }
   \left( \Delta u
        , \Delta P
        , \Theta
        , \hslash \right) \simeq
   \Zcal_{ F_{ m - 1 } }
   \left( \delta \, \Delta u
        , \eta \, \Delta P
        , \Jcal \, \Theta
        , \kappa \, \hslash \right)
   \left( \frac{ \sqrt{\Jcal}
              }{ \left| \alpha \right| \varepsilon }
   \right)^{ J F_{ m - 1 } } .
\ee
\begin{figure}[!ht]
\centering
   \psfrag{T}[cb][][0.8][0]{$\Theta$}
   \psfrag{h}[cc][][0.8][0]{$\hslash$}
   \psfrag{cP}[cb][][1.0][0]{$c_\Psss \left( \Theta , \hslash \right)$}
   \includegraphics[width=0.5\textwidth]{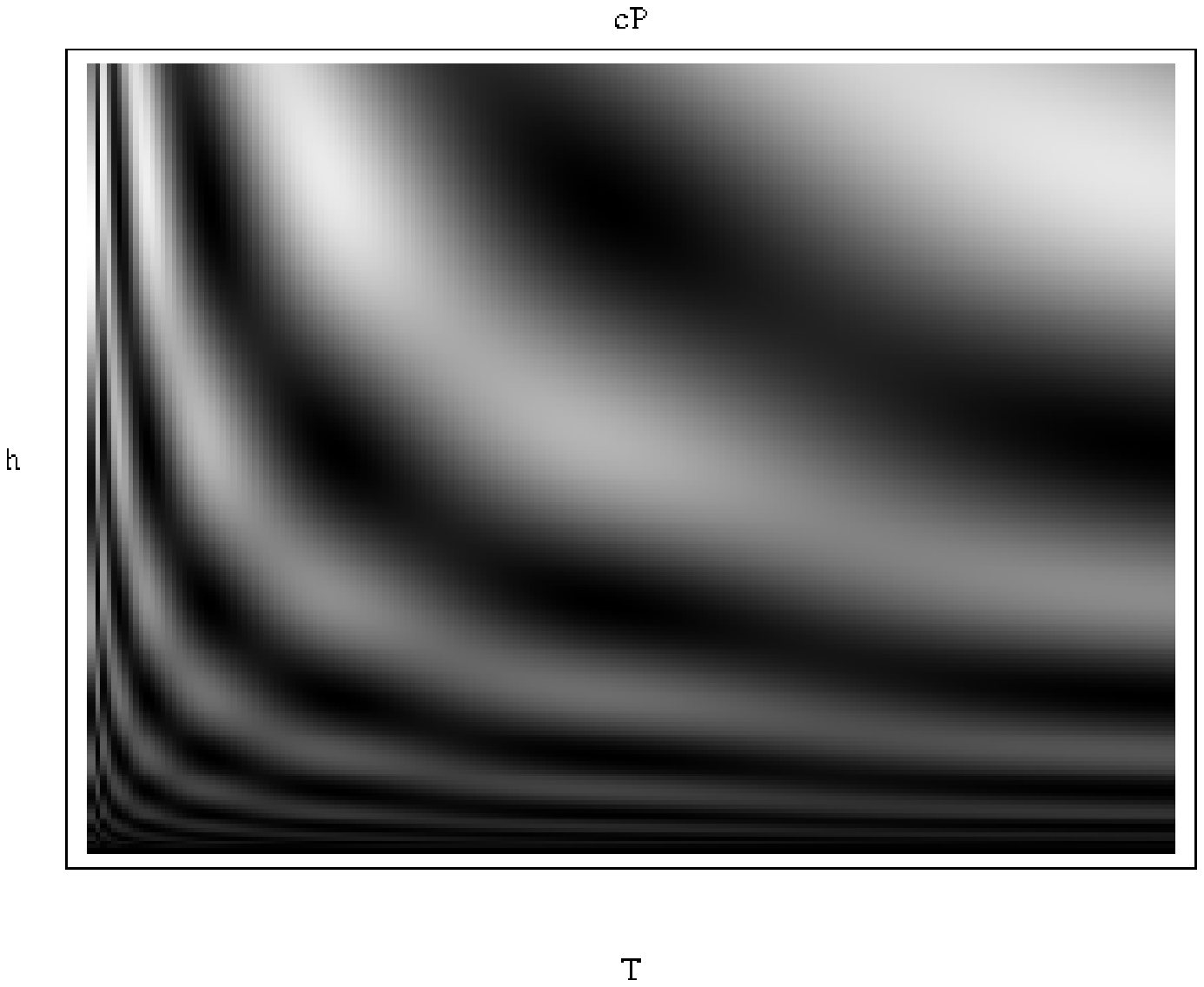}
   \caption{\small Example of the behaviour of $c_\Psss$ for the
choice of $k \left( a , b \right) =
\left[ c^{ \sss + } k^{ \sss + } \left( a , b \right)
     + c^{ \sss - } k^{ \sss - } \left( a , b \right) \right]
\erm^{ - c \left( \frac{a}{ \ln \Jcal }
                - \frac{b}{ \ln \kappa } \right) }$ in
equation~(\ref{eq:cPEquation_Q}) with
$c = 1 / 2$, $c^{ \sss + } = 1$, $c^{ \sss - } = 1 / 5 $
and $k^\pmsss = 1 + \cos 2 \pi \left( \frac{a}{ \ln \Jcal }
\pm \frac{b}{ \ln \kappa } \right)$.}
\label{fig:schottky_Q}
\end{figure}
The free energy $f = - \Theta \lim_{ m \rightarrow \infty}
\ln \Zcal_{ F_m } / F_m$ therefore behaves like (using
$\gamma^{ - 1 }
= \lim_{ m \rightarrow \infty } F_{ m - 1 } / F_m $)
\be
\nonumber
   f \left( \Delta u
          , \Delta P
          , \Theta
          , \hslash \right)
   \simeq \frac{1}{ \gamma \Jcal }
   f \left( \delta \, \Delta u
          , \eta \, \Delta P
          , \Jcal \, \Theta
          , \kappa \, \hslash \right)
   - \frac{ \Theta J }{\gamma}
   \ln \left( \frac{ \sqrt{\Jcal}
                  }{ \left| \alpha \right| \varepsilon } \right) ,
\ee
close to the TBA fixed point, and
$ e = - \Theta^2 \, \p \left( f / \Theta \right) / \p \Theta$,
the energy per particle, like

\be
\nonumber
   e \left( \Delta u
          , \Delta P
          , \Theta
          , \hslash \right)
\nonumber
   \simeq \frac{1}{ \gamma \Jcal }
        e \left( \delta \, \Delta u
               , \eta \, \Delta P
               , \Jcal \, \Theta
               , \kappa \, \hslash \right) .
\ee
This leads to the following scaling law for the specific heat
per particle at constant pressure $c_{ \Psss } = \p e / \p \Theta$:
\be
\label{eq:cPScaling_Q}
   c_\Psss \left( \Delta u
                , \Delta P
                , \Theta
                , \hslash \right)
   \simeq \gamma^{ - 1 }
   c_\Psss \left( \delta \, \Delta u
          , \eta \, \Delta P
          , \Jcal \, \Theta
          , \kappa \, \hslash \right) .
\ee
In particular close to the TBA, with $\Delta u =\Delta P = 0$
for small $\Theta$ and $\hslash$, equation (\ref{eq:cPScaling_Q})
suggests that in the $\left( \Theta , \hslash \right)$-plane the
specific heat shows a sequence of modulated ridges, which are
invariant under scaling by $\left( \Jcal , \kappa \right)$:
\be
\label{eq:cPEquation_Q}
   c_\Psss \left( \Theta , \hslash \right) \simeq
   \Theta^{ \frac{ \ln \gamma }{ \ln \Jcal } } \,
   k \left( \ln \Theta , \ln \hslash \right) ,
\ee
with
$k \left( a + \ln \Jcal , b + \ln \kappa \right)
= k \left( a , b \right)$,
for all $a$ and $b$ (see figure~\ref{fig:schottky_Q}).
These ridges thus constitute the extension to the
$\left( \Theta , \hslash \right)$-plane of the sequence of
{\em Schottky anomalies} responsible for the phenomenon of
{\em hierarchical melting} of the chain
\cite{art:ValletEtAl86,art:SchillingAubry87,art:ValletEtAl88}.


\section*{Acknowledgements}
NRC was funded by the Portuguese institution `Funda\c{c}\~{a}o
para a Ci\^{e}ncia e Tecnologia' under the reference
PRAXIS~XXI/BD/21987/99.

\bibliography{bibliography}

\end{document}